Solving Schrödinger Equation for Three-Electron Quantum Systems by the Use of The Hyperspherical Function Method

Lia Leon Margolin [1], Shalva Tsiklauri [2]

[1] *Marymount Manhattan College, New York, NY*
[2] *New York City College of Technology, CUNY, Brooklyn, NY,*

**Abstract**

A new model-independent approach for the description of three electron quantum dots in two dimensional space is developed. The Schrödinger equation for three electrons interacting by the logarithmic potential is solved by the use of the Hyperspherical Function Method (HFM). Wave functions are expanded in a complete set of three body hyperspherical functions. The center of mass of the system and relative motion of electrons are separated. Good convergence for the ground state energy in the number of included harmonics is obtained.

**1. Introduction**

One of the outstanding achievements of nanotechnology is construction of artificial atoms- a few-electron quantum dots in semiconductor materials. Quantum dots [1], artificial electron systems realizable in modern semiconductor structures, are ideal physical objects for studying effects of electron-electron correlations. Quantum dots may contain a few two-dimensional (2D) electrons moving in the plane z=0 in a lateral confinement potential V(x, y).

Detailed theoretical study of physical properties of quantum-dot atoms, including the Fermi-liquid-Wigner molecule crossover in the ground state with growing strength of intra-dot Coulomb interaction attracted increasing interest [2-4]. Three and four electron dots have been studied by different variational methods [4-6] and some important results for the energy of states have been reported. In [7] quantum-dot Beryllium (N=4) as four Coulomb-interacting two dimensional electrons in a parabolic confinement was investigated. Energy spectra, charge and spin densities, and electron-electron correlations in a harmonic oscillator potential were obtained by the use of Exact Diagonal Approximation method. However, all above mentioned methods are applicable only if confinement energy is much larger than electron-electron interaction energy. In lateral quantum dots, defined by metallic gates in a 2D electron gas confinement energy and electron-electron interaction energy are almost the same order, therefore above listed approximate methods cannot provide adequate description of the system.

Theoretical study of the physical properties of quantum dots as a function of external magnetic field is a very important problem to solve since it will allows us to tune physical properties of these dots by experimentally changing external magnetic field frequency.

In the next two sections a new model-independent approach for the description of three electron quantum dots in 2D space is developed by the use of the Hyperspherical Function Method (HFM) presented in [8-10]. The wave functions of three electron system are expanded in a complete set of three body hyperspherical functions, and potential energy of the interaction between confined electrons is described by the logarithmic function. The HFM allows us to separate the center of mass and relative motion of electrons and obtain ground state energies as a function of an external magnetic field by solving Schrödinger's equation.

## 2. Mathematical Modeling of Three Electron Quantum Dots in 2 D Space by the Use of The Hyperspherical Function Method

Theoretical studies of three electron quantum dots have been carried out only for Coloumb interacting electrons. However, due to the fact that the solution of Poisson equation in 2 D space for three electron quantum system is represented by the logarithmic function, it is extremely important to describe electron-electron interactions with logarithmic potential.

Solving Schrodinger equation for two dimensional electrons in a parabolic confinement with Hyperspherical Function Method (HFM) allows us to separate the center of mass movement and consider logarithmic potential of electron -electron interactions.

Hamiltonian for three Electrons in parabolic confinement can be written as:

$$H = \sum_{i=1}^{3}\left[\frac{1}{2m_{eff}}\left(\vec{p}_i - \frac{e}{c}\vec{A}_i\right)^2 + \frac{1}{2}m^*\omega_0^2 r_i^2\right] + \sum_{i \neq j}^{3} V(\vec{r}_{ij}) \quad (1)$$

Where $\vec{p}_i$ is the generalized momentum of the i-th particle, $\vec{A}_i$ vector potential of the magnetic field at the point occupied by i-th particle

$$\vec{A}_j = \frac{1}{2}\vec{B} \times \vec{r}_i = \frac{1}{2}B(-y_i, x_i, 0)$$

$m_{eff}$ is an effective mass of an electron, $\omega_0$ is strength of confinement, $r_0 = \sqrt{\frac{\hbar\omega_0}{m^*}}$ unit of length, and V(r) is electron-electron interaction potential. If we substitute logarithmic potential in (1.1) we will obtain the following expression:

$$H = \sum_{i=1}^{3}\left(-\frac{\hbar^2}{2m_{eff}}\nabla_i^2 + \frac{m^*\omega^2}{2}r_i^2\right) - \sum_{i \neq j}^{3}\beta\ln\frac{|\vec{r}_{ij}|}{r_0} - \omega_L L_z$$

Where $\omega_L = eB/2m^*c$ is Larmor frequency, and $\omega = (\omega_0^2 + \omega_L^2)^{1/2}$

Let's introduce mass-scaled Jacobi coordinates $\hat{\vec{X}}_i, \hat{\vec{Y}}_j$, and R of three particle systems defined by:

$$\vec{X}_i = \left(\frac{m_j m_k}{(m_j + m_k)\mu}\right)^{1/2} (\vec{r}_j - \vec{r}_k)$$

$$\vec{Y}_i = \left(\frac{m_j(m_j + m_k)}{M\mu}\right)^{1/2} \left(-\vec{r}_j + \frac{m_j \vec{r}_j + m_k \vec{r}_k}{m_j + m_k}\right) \quad (2)$$

$$\vec{R} = \left(\frac{1}{\sqrt{M\mu}}\right) \sum_{i=1}^{3} m_i r_i$$

Where $M = m_1 + m_2 + m_3$ $\{i,j,k,\} = \{1,2,3\}$, and $\mu = \left(\frac{m_1 m_2 m_3}{M}\right)^{1/2}$ is a reduced mass of three particle system. For the systems of three identical particles (electrons) $m_1 = m_2 = m_3 = m$, $\mu = \frac{m}{\sqrt{3}}$ and Jacobi coordinates can be found easily.

Schrodinger equation with Hamiltonian (1) can be written as:

$$\left(-\frac{1}{4}\nabla_R^2 + \omega^2 R^2 - \frac{\hbar^2}{2\mu}(\Delta_{\vec{X}} + \Delta_{\vec{Y}}) + \frac{1}{4}\omega^2(X^2 + Y^2) + U_{123}(\vec{X},\vec{Y})\right)\Psi(1,2,3) = E\Psi(1,2,3) \quad (3)$$

This equation enables us to represent Eigenfunctions of three electrons as a following product:

$$\psi(1,2,3) = \varphi(x,y)\phi(R)\sigma(s_1,s_2,s_3) \quad (4)$$

Where $\sigma(s_1,s_2,s_3)$ are spin functions identifying parity of both $\phi(R)$ and $\varphi(x,y)$ functions, $\varphi(x,y)$ describes the relative motion of electrons, $\phi(R)$ describes the movement of the center of mass. If we substitute (4) into (3) we will receive the equation of the linear harmonic oscillator for $\phi(R)$.

$$\left(-\frac{1}{4}\nabla_R^2 + \omega^2 R^2\right)\phi(\vec{R}) = E_R \phi(\vec{R}) \quad (5)$$

Where $E = E_R + E_{x,y}$ and $\omega_L = eB/2mc$ is Larmor frequency and $L = l_x + l_y$. Hamiltonian in the equation (5) coincides with one electron Hamiltonian in 2D parabolic confinement and gives us the following Fock-Darvin energy levels (See Fig. 1):

$$E_{n,m}^{CM} = \hbar\omega(2n + |m| + 1) - \hbar\omega_L m \quad (7)$$

As for the relative movements of free electrons $\varphi(x,y)$ if we substitute (1.5) into (1.3) we will obtain the following equation

$$\left(-\frac{\hbar^2}{2\mu}(\Delta_{\vec{x}}+\Delta_{\vec{y}})+\tfrac{1}{4}\omega^2(x^2+y^2)+U_{123}(\vec{x},\vec{y})-\omega_L L_z^{tot}\right)\varphi(\vec{x},\vec{y})=E_{x,y}\varphi(\vec{x},\vec{y}) \quad (8)$$

In order to solve equation (1.8) for the relative motion of the electrons let's introduce the Hyperspherical coordinates in Four dimensional Euclidean space.

$$-\infty \leq \rho \leq +\infty\,;\ \vec{n}_{x_i}=\frac{\vec{X}_i}{|\vec{X}_i|}\,;\ \vec{n}_{y_i}=\frac{\vec{Y}_i}{|\vec{Y}_i|}\,;\ 0\leq\alpha\leq\pi/2\,;\ \{\rho,\Omega_i\}=\{\rho,\alpha_i,\hat{\vec{x}}_i,\hat{\vec{y}}_i\}$$

$$\rho^2 = X_1^2+Y_1^2 = X_2^2+Y_2^2 = X_3^2+Y_3^2\,;\ |\vec{X}_i|=\rho\cos\alpha\,;\ |\vec{Y}_i|=\rho\sin\alpha \quad (9)$$

where $\hat{\vec{x}}_i, \hat{\vec{y}}_i$ define directions of $\vec{X}_i$ and $\vec{Y}_i$ vectors. Relationship between Hyperspherical coordinates can be written in expanded way:

$$\begin{aligned}
|\vec{X}_{x1}| &= \rho\cos\alpha\cos\varphi_1; & |\vec{Y}_1| &= \rho\sin\alpha\cos\varphi_2;\\
|\vec{X}_2| &= \rho\cos\alpha\sin\varphi_1; & |\vec{Y}_2| &= \rho\sin\alpha\sin\varphi_2; \quad (10)\\
|\vec{X}_3| &= \rho\cos\alpha; & |\vec{Y}_3| &= \rho\sin\alpha.
\end{aligned}$$

Connection between different sets of Jacobi coordinates (2) can be represented as:

$$\vec{X}_k = \vec{X}_i\cos\phi_{ik}+\vec{Y}_i\sin\phi_{ik}$$
$$\vec{y}_k = -\vec{X}_i\sin\phi_{ik}+\vec{Y}_i\cos\phi_{ik} \quad (11)$$

The angle $\phi_{ik}$ can be easily found using the following formula.

$$\phi_{ik} = \arctan\left[(-1)^p\left(\frac{m_j M}{m_k m_i}\right)^{1/2}\right] \quad (12)$$

Where p can be either odd or even depending on the parity of the $\{i,k,j\}$ particle permutations. Obtaining $\phi_{ik}$ from (12) isn't difficult for the systems of three identical particles.

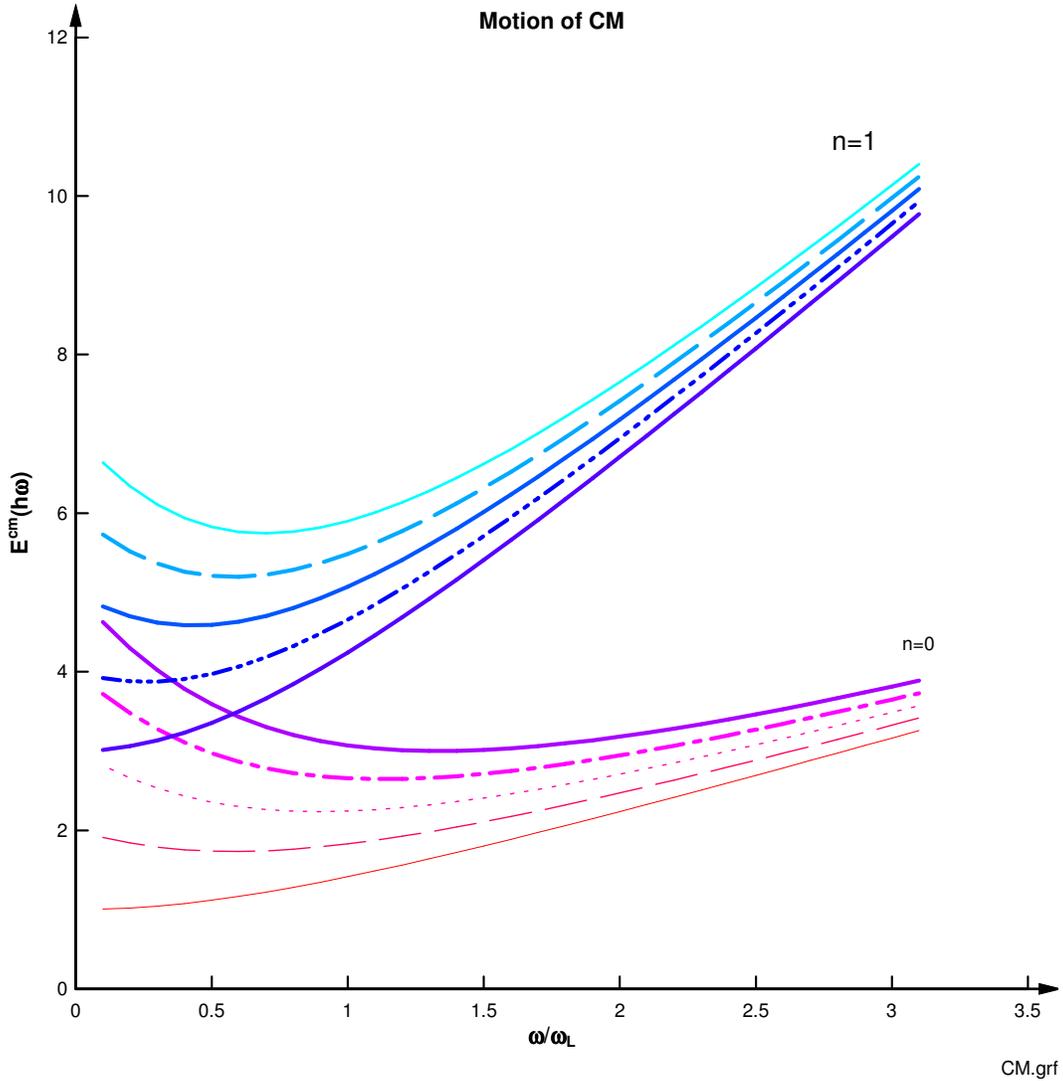

Fig 1.1 a) Energy levels of the movement of the center of mass as a function of $\omega_L/\omega$ when n=0,1 and m=0,1,...5 Dotted lines represent Landau levels

Kinetic energy operator for three particle systems in Hyperspherical coordinates can be represented as:

$$T = \frac{\partial^2}{\partial \rho^2} + \frac{3}{\rho}\frac{\partial}{\partial \rho} - \frac{K^2(\Omega_i)}{\rho^2} \quad (13)$$

$K^2(\Omega_i)$ is the square of the four-dimensional space angular momentum

$$K^2(\Omega_i) = +\frac{\partial^2}{\partial \alpha^2} + 2ctg2\alpha\frac{\partial}{\partial \alpha} - \left(\frac{1}{\cos^2 a}l_1^2 + \frac{1}{\sin^2 \alpha}l_2^2\right) \quad (14)$$

where $l_i^2 = \dfrac{\partial^2}{\partial \varphi_j^2}$ (i=1,2) represent squares of the impulses of the corresponding Jacobi vectors

Let's rewrite equation (6) for the relative motion of free electrons in Hyperspherical coordinates:

$$\left( \frac{\partial^2}{\partial \rho^2} + \frac{3}{\rho}\frac{\partial}{\partial \rho} - \frac{K^2(\Omega_i)}{\rho^2} - \frac{2\mu}{\hbar^2}U_{123}(\vec{x},\vec{y}) - \tfrac{1}{4}\omega^2 \rho^2 + \frac{2\mu}{\hbar^2}E - \omega_L L_Z^{rel} \right)\psi(\rho,\alpha) = 0 \quad (15)$$

Where $U_{123}(\vec{x},\vec{y})$ represents potential energy of interacting particles. $K^2(\Omega_i)$ is an angular part of four dimensional Laplace operator moment with the eigenstates of $K(K+1)$, and eigenfunctions that create complete set of the basic ortonormal Hyperspherical functions:

$$\Phi_{KLM}^{l_1 l_2}(\Omega) = N_{l_1 l_2}^n P_n^{l_1 l_2}(\alpha) e^{i l_1 \varphi_1} e^{i l_2 \varphi_2} \quad (16)$$

Where

$$P_n^{l_1 l_2}(\alpha) = \sum_{m=0}^{\nu}(-1)^{\nu - m}\binom{\nu + |l_2|}{m}\binom{\nu + |l_1|}{\nu - m}(\cos\alpha)^{2m + l_1}(\sin\alpha)^{2(\nu - m) + |l_2|}$$

$$L = \hbar(l_1 + l_2)$$

$$K = 2n + |l_1| + |l_1|, \quad N = \sqrt{(2n + |l_1| + |l_2| + 1)n!(n + |l_1| + |l_2|)! / [2\pi^2 (n + |l_1|)!(n + |l_2|)!]}$$

Let's expand $\Psi(\rho,\alpha)$ wave functions in Hyperspherical basis:

$$\psi(\rho,\alpha) = \sum_{K l_1 l_2 M} \rho^{-3/2} \chi_{KL}^{l_1 l_2}(\rho) \Phi_{KLM}^{l_1 l_2}(\Omega_i) \quad (17)$$

after substitution of (17) into (15) for hyper radial wave functions we will receive infinite system of the following equations:

$$\left( \frac{\partial^2}{\partial \rho^2} - \left[ \chi^2 + \frac{(K+1)^2 - 0{,}25}{\rho^2} \right] \right)\varphi_{RL}^{l_1 l_2}(\rho) = \sum_{K' l'_1 l'_2 M'} W_{KK'LL'MM'}^{l_1 l_2 l'_1 l'_2}(\rho)\varphi_{K'L'}^{l'_1 l'_2}(\rho) \quad (18)$$

Where:

$$W_{KK'LL'MM'}^{l_1 l_2 l'_1 l'_2}(\rho) = \frac{2\mu}{\hbar^2}\int \Phi_{KLM}^{l_1 l_2}(\Omega_i) U_{123}(\alpha)\Phi_{K'L'M'}^{l'_1 l'_2}(\Omega_i) d\Omega_i \quad (19)$$

and $\chi^2 = -\dfrac{2\mu}{\hbar^2}E$ (20)

In order to calculate overlapping integral between have functions defined on the different sets of Jacobi coordinates we have to use Reinal-Revai unitary transformation coefficients defined by the formula;

$$\Phi_{KLM}^{l_1 l_2}(\Omega_i) = \sum_{\tilde{l}_1 \tilde{l}_2} {}^i\langle \tilde{l}_1 \tilde{l}_2 | l_1 l_2 \rangle_{KL}^j \Phi_{KLM}^{\tilde{l}_1 \tilde{l}_2}(\Omega_i)^i \quad (21)$$

Where $\langle \tilde{l}_1\tilde{l}_2 | l_1l_2 \rangle^j_{KL}$ represent unitary transformation coefficients first introduced by Reinal and Revai [11].

Taking into consideration (21), overlapping integral (19) can be rewritten in a following way:

$$W^{l_1l_2l'_1l'_2}_{KK'LL'MM'}(\rho) = \frac{2\mu}{\hbar^2}\left[ J^{l_1l_2l'_1l'_2}_{KK'LL'MM'} + \sum_{\tilde{l}_1\tilde{l}_2} {}^i\langle \tilde{l}_1\tilde{l}_2 | l_1l_2 \rangle^j_{KL} {}^i\langle \tilde{l}'_1\tilde{l}'_2 | l'_1l'_2 \rangle^j_{KL} J^{\tilde{l}_1\tilde{l}_2\tilde{l}'_1\tilde{l}'_2}_{KK'LL'MM'} + \right.$$

$$\left. + \sum_{\tilde{l}_1\tilde{l}_2} {}^i\langle \tilde{\tilde{l}}_1\tilde{\tilde{l}}_2 | l_1l_2 \rangle^j_{kL} {}^i\langle \tilde{\tilde{l}}'_1\tilde{\tilde{l}}'_2 | l'_1l'_2 \rangle^j_{KL} J^{\tilde{\tilde{l}}_1\tilde{\tilde{l}}_2\tilde{\tilde{l}}'_1\tilde{\tilde{l}}'_2}_{KK'LL'MM'} \right] \quad (22)$$

Where angular integral for instance between first and second particles can be written as:

$$J^{l_1l_2l'_1l'_2}_{KK'LL'MM'} = \int \Phi^{*l_1l_2}_{KLM}(\Omega_i) U_{12} \Phi^{l'_1l'_2}_{K'L'M'}(\Omega_i) d\Omega \quad (23)$$

$\Phi^{l_1l_2}_{KLM}(\Omega)$ can be found from (15) and $U_{12}$ is the part of the potential energy corresponding interaction between first and second particles.

When considering system of identical particles Hyperspherical functions in (1.17) need to be replaced with symmetrized three body Hyperspherical functions.
For three electron systems we will obtain the following expansion:

$$\psi(\rho,\alpha) = \sum_{K[f]\nu_{[f]}M} \rho^{-3/2} \varphi^{[f]\nu_f}_{KL}(\rho) \Gamma^{[f]\nu_f}_{KLM}(\Omega,\vec{\sigma}) \quad (24)$$

Where

$$\Gamma^{[f]\nu_f}_{KLM}(\Omega,\vec{\sigma}) = \frac{1}{h_{[f]}} \sum_{\mu_{[f]}} \Phi^{[f]\nu_{[f]}}_{KLM}(\Omega) \chi^{[\tilde{f}]\tilde{\mu}_{\tilde{f}}}(\vec{\sigma}) \quad (25)$$

[f] is the Young diagram of the three particle system, $h_{[f]}$ indicates dimension of [f] representation, $\mu_{[f]}$ denotes the rows of the [f] representation, and $\nu_{[f]}$ is the [f] representation number with given K and L, $\Phi^{[f]\nu_{[f]}}_{KLM}$ are symmetrized Hyperspherical functions defined by the following expansion:

$$\Phi^{[f]\nu_{[f]}}_{KLM}(\Omega) = \sum_{l_1l_2} C^{[f]\nu_{[f]}}_{KL}(\Omega) \Phi^{l_1l_2}_{KL}(\Omega) \quad (26)$$

Three body symmetrization coefficients $C^{[f]\nu_{[f]}}_{KL}$ can be obtained using three body Reinal-Revai coefficients. [8-9]

## 3. Solving Three Electron Schrodinger Equation with Logarithmic Potential

Logarithmic potential of electron-electron interactions is given with the following formula:

$$V_{ij}(\vec{r}_{ij}) = \frac{e_i e_j}{\varepsilon} \ln \frac{|\vec{r}_{ij}|}{r_0} \quad (27)$$

If we substitute (27) in (23) for the overlapping integral of the Hyperspherical functions we will obtain [12]:

$$J_{KK'LL'MM'}^{l_1 l_2 l_1' l_2'} = N_{l_1' l_2'}^{n'} N_{l_1 l_2}^{n} \sum_{m'=0}^{v'} \sum_{m=0}^{v} (-1)^{v'-m'} (-1)^{v-m} \binom{v'+|l_2'|}{m'} \binom{v'+|l_1'|}{v-m'} \binom{v+|l_2|}{m} \binom{v'+|l_1|}{v-m} \int (\cos\alpha)^{2m+|l_1|+2m'+|l_1'|} *$$

$$(\sin\alpha)^{2(v-m)+|l_1|+2(v'-m')+|l_2'|} \ln(\rho \cos\alpha) \cos(\alpha) \sin(\alpha) d\alpha =$$

$$= \delta_{n'n} \delta_{l_1 l_1'} \delta_{l_2 l_2'} \delta_{LL'} \delta_{mm'} e_i e_j \frac{1}{\varepsilon} \ln\left(\frac{\rho}{\rho_0}\right) + e_i e_j \frac{1}{\varepsilon} C_{[L][L']} \quad (28)$$

where

$$C_{LL'} = N_{l_1' l_2'}^{n'} N_{l_1 l_2}^{n} \sum_{m'=0}^{v'} \sum_{m=0}^{v} (-1)^{v'-m'} (-1)^{v-m} \binom{v'+|l_2'|}{m'} \binom{v'+|l_1'|}{v-m'} \binom{v+|l_2|}{m} \binom{v'+|l_1|}{v-m} \int (\cos\alpha)^{2m+|l_1|+2m'+|l_1'|} \times$$

$$\times \int (\cos\alpha)^{2m+|l_1|+2m'+|l_1'|} (\sin\alpha)^{2(v-m)+|l_1|+2(v'-m')+|l_2'|} \ln(\cos\alpha) \cos\alpha \sin\alpha d\alpha \quad (29)$$

Taking into consideration (28) the system of equations (18) can be rewritten in the following way:

$$\left(\frac{\partial^2}{\partial \rho^2} - \left[\chi^2 + \frac{(K+1)^2 - 0,25}{\rho^2} + \frac{1}{4}\omega^2 \rho^2\right]\right)\varphi_{KL}^{l_1 l_2}(\rho) =$$

$$= \sum_{K' l_1' l_2' M'} \left(\delta_{n'n} \delta_{l_1 l_1'} \delta_{l_2 l_2'} \delta_{LL'} \delta_{mm'} e_i e_j \frac{1}{\varepsilon} \ln\left(\frac{\rho}{\rho_0}\right) + C_{[L][L']}\right) \varphi_{K'L'}^{l_1' l_2'}(\rho) \quad (30)$$

It is important to note that coefficients of $\ln\left(\frac{\rho}{\rho_0}\right)$ are diagonal and coefficients of all other terms are independent from $\rho$. This is what distinguishes logarithmic potential from all other potentials. In this article we will only consider diagonal terms. This approach is justified by the convergence of the Hyperspherical expansion.

The Diagonal part of the equation (30) is simple:

$$\left(\frac{\partial^2}{\partial \rho^2} + \chi^2 - \frac{(K+1)^2 - 0.25}{\rho^2} - \frac{1}{4}\omega^2 \rho^2 - \varepsilon_i \varepsilon_j \frac{1}{\varepsilon} \ln\left(\frac{\rho}{\rho_0}\right) + C_{LL}\right)\varphi_{kl}(\rho) = 0 \quad (31)$$

Let's consider ground state L=0, $K_{min} \geq L$, where $L = l_1 + l_2$ than from (31) system of equations we will only have one equation left:

$$\left( \frac{\partial^2}{\partial \rho^2} + \chi^2 - \frac{(K+1)^2 - 0.25}{\rho^2} - \frac{1}{4}\omega^2 \rho^2 - \varepsilon_i \varepsilon_j \frac{1}{\varepsilon} \ln\left(\frac{\rho}{\rho_0}\right) + e_i e_j \frac{1}{2\pi^2 \varepsilon} \right) \varphi_{kl}(\rho) = 0 \quad (32)$$

After solving (32) when $m^* = 0.067m$, $\varepsilon_r = 12$ (GaAs), $\hbar\omega = 5\, mev$

We will obtain the following wave functions and energy levels:

$$\varphi_{KN}^\rho(\rho) = \sqrt{2} \frac{N!}{(N+K+1)!} (\alpha\rho^2)^{K+1} \exp\left(-\alpha\rho^2/2\right) L_N^{K+1}(\alpha\rho^2); \quad (33)$$

$$E^0 = \sqrt{\frac{2}{3}} \hbar\omega(2N + K + 2) - \hbar\omega_M L_z \quad (34)$$

Where: N=0,1,…

$$\alpha = \sqrt{\frac{2m\omega^2}{3\hbar}}$$

If external magnetic field is weak, than we have (L,S)=(0, ½) configuration corresponding to symmetric radial wave functions and we will obtain the following energy levels (Fig. 2a). If external magnetic field is strong, than we have completely polarized state S=3/2 and (L,S)=(1, 3/2) configuration corresponding to antisymmetric radial wave functions and the energy levels obtained from (2.8) coincide with Landau energy levels. (Fig. 2b).

I order to solve equation (30) of relative motion of electrons in non-diagonal approximation, let's expand wave function into complete set of basic Hyperradial functions.

$$\varphi_K^{l_1 l_2}(\rho) = \sum_{N=0}^{\infty} a_{KN} \varphi_{KN}^{0 l_1 l_2}(\rho) \quad (35)$$

Coefficients of this expansion obey normalization conditions

$$\sum_{N=0}^{\infty} \left| a_{KN}^{l_1 l_2} \right|^2 = 1 \quad (36)$$

If we substitute (35) in (30) the system of differential equations will be reduced to the infinite system of linear homogenous algebraic equations and the energy eigenvalues can be obtained from the requirement that the determinant

$$det \left\| (E - E_0^{K'N'}) \delta_{KK'} \delta_{l_1 l_1'} \delta_{l_2 l_2'} \delta_{NN'} - I_{KK'LL'MM'}^{l_1 l_1' l_2 l_2'} (1 - \delta_{KK'} \delta_{l_1 l_1'} \delta_{l_2 l_2'} \delta_{NN'}) \right\| = 0 \quad (37)$$

Where:

$$I = 2\alpha^2 \left(\frac{n!}{(n+k+1)!}\right)^2 \int_0^\infty (\alpha\rho^2)^{2(K+1)} \exp(\alpha\rho^2) L_N^{K+1}(\alpha\rho^2) \ln(\rho/\rho_0) d\rho \qquad (38)$$

This integral can be solved analytically.

System of equations (36) has been solved for K=0,2,4,6. Energy levels for interacting electrons are presented at fig 3. Good convergence (energy levels for K=4 and K=6 are almost the same) to the value 0.284 mev for the ground state energy in the number of included harmonics is obtained.

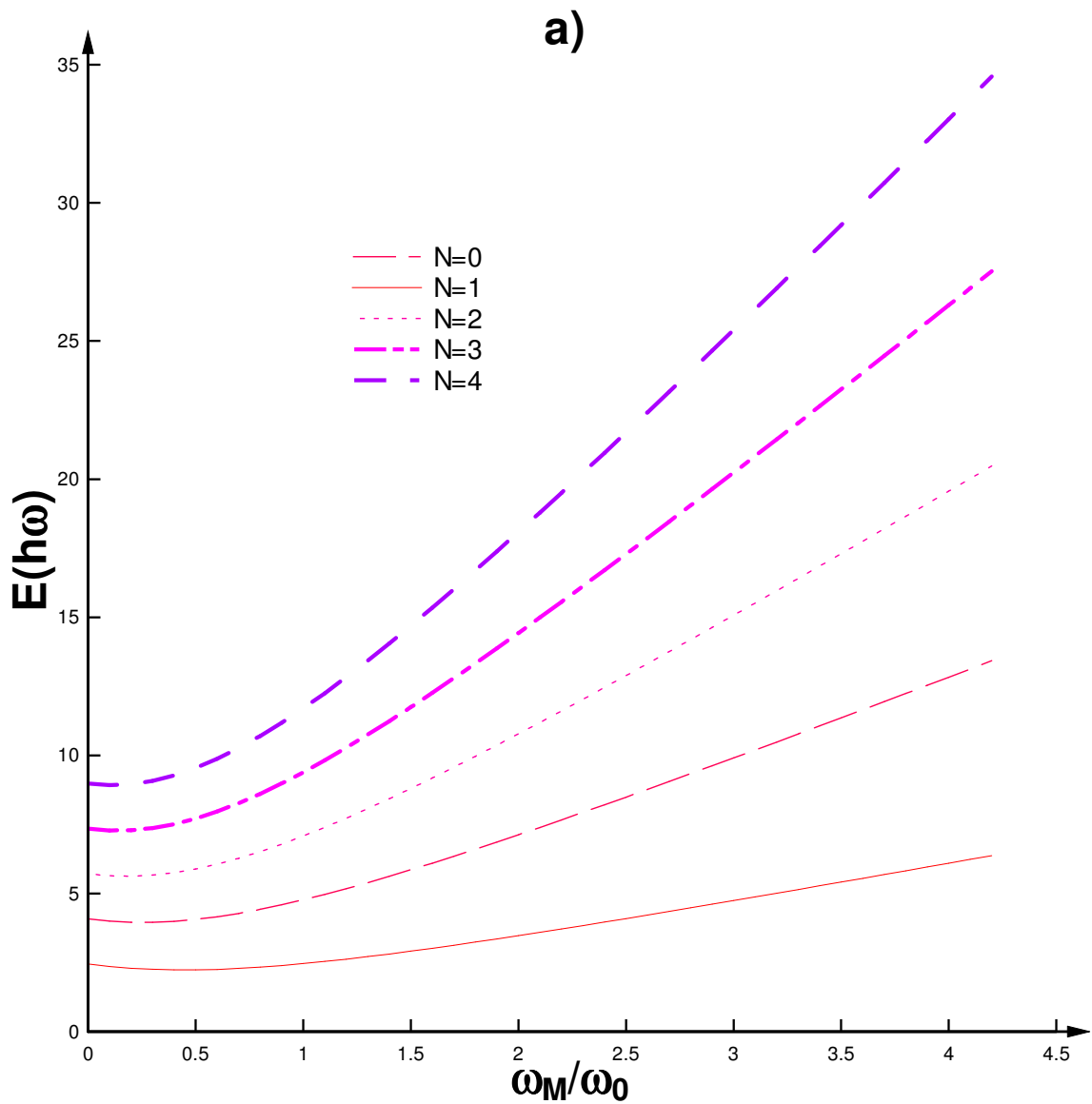

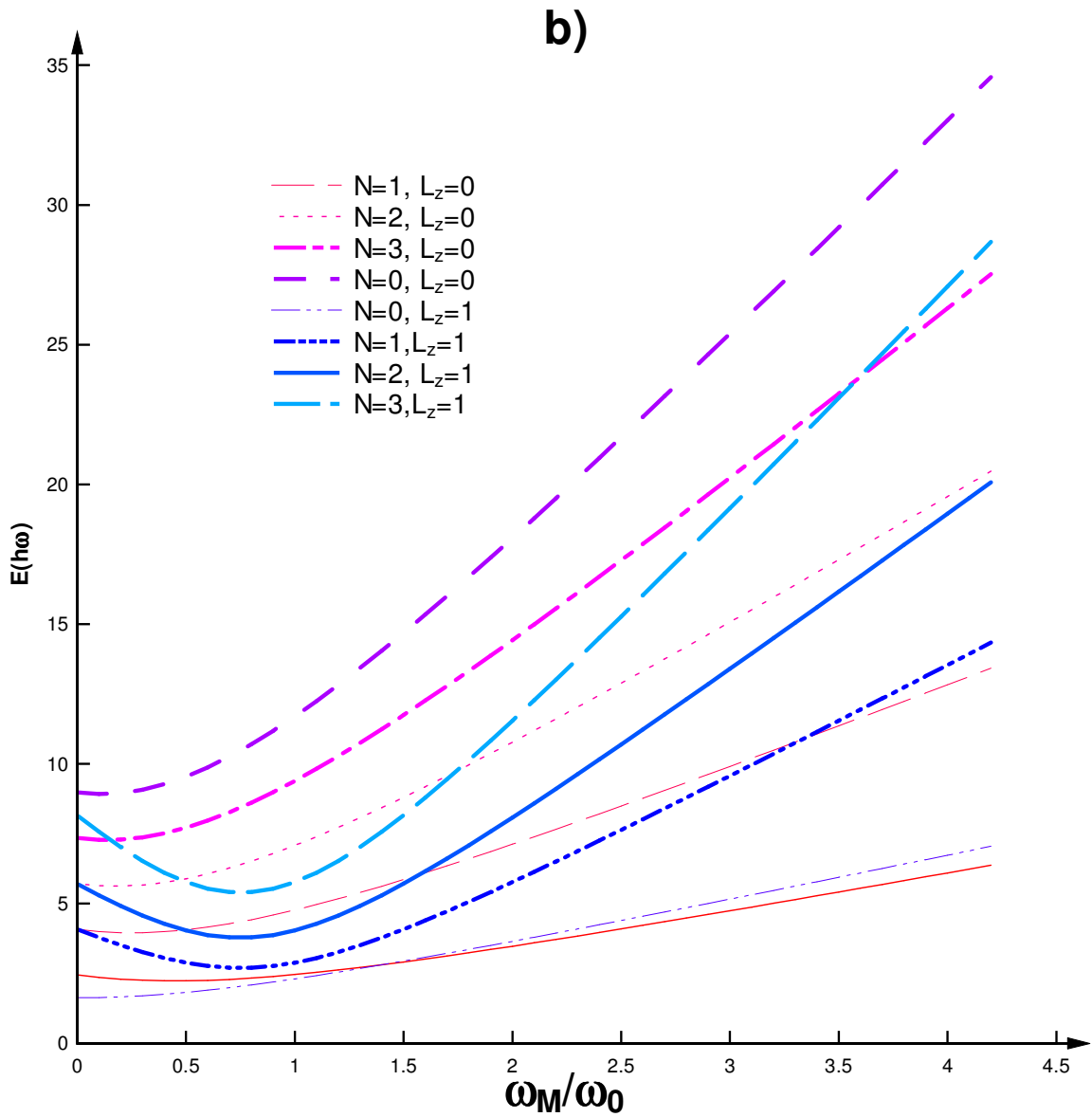

Fig 2. Energy levels of the relative motion of non-interacting electrons for:
a) K=L=0 , N=0,1,2,3, 4 and b) K=L=1, N=0,1,2,3

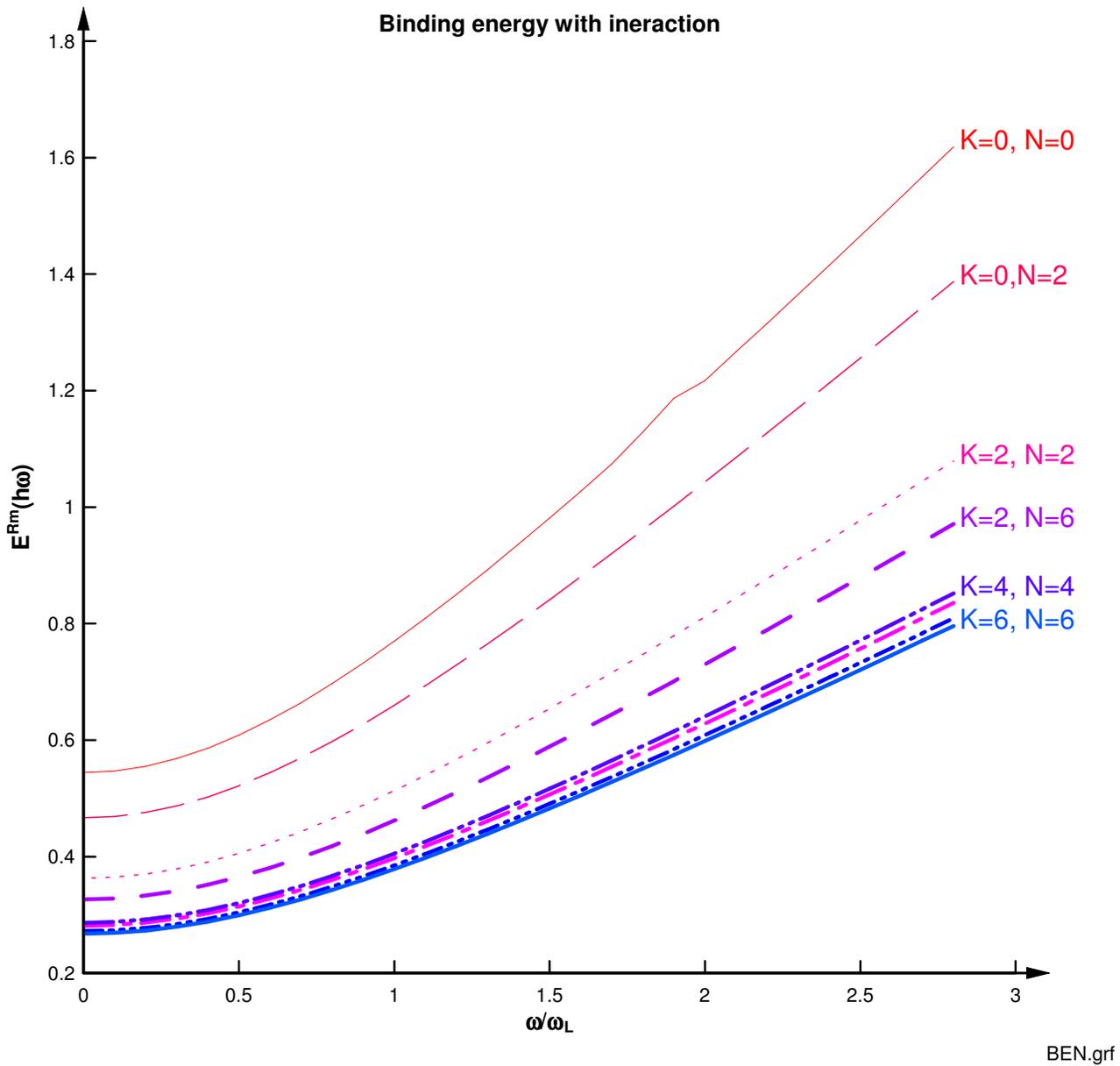

Fig 3 Energy levels of interacting electrons when K=0,2,4,6

## 4. Conclusion.

Method of Hyperspherical functions proved to be very effective for the investigation of three electron quantum dots in 2 D space. First time three electron quantum dots have been studied by the use of logarithmic electron-electron potential. Obtained theoretical results for the ground state energies demonstrated satisfactory agreement with existing experimental data.